\documentclass{aa}
\usepackage{graphics}
\def\ltsima{$\; \buildrel < \over \sim \;$}
\def\simlt{\lower.5ex\hbox{\ltsima}}    
\def\gtsima{$\; \buildrel > \over \sim \;$}
\def\simgt{\lower.5ex\hbox{\gtsima}}    

\def\ref{\par\noindent\hangindent 20 pt}

\def\mincir{\ \raise -2.truept\hbox{\rlap{\hbox{$\sim$}}\raise5.truept 
\hbox{$<$}\ }}  %
\def\magcir{\ \raise -2.truept\hbox{\rlap{\hbox{$\sim$}}\raise5.truept %

\hbox{$>$}\ }}

\def\approxlt{\mathrel{\hbox{ \lower .5ex \hbox {$\sim$}
	\llap{\raise .15 ex \hbox{$<$}} }}}
\def\approxgt{\mathrel{\hbox{ \lower .5ex \hbox {$\sim$}
	\llap{\raise .15 ex \hbox{$>$}} }}}

\def\multleft#1{\hbox to size{\vbox {\halign {\lft{##}\cr #1}}\hfill}\par}
\def\multright#1{\hbox to size{\vbox {\halign {\rt{##}\cr #1}}\hfill}\par}

\def\today{\ifcase\month\or January\or February\or March\or April\or May\or
      June\or July\or August\or September\or October\or November\or December\fi
      \space\number\day, \number\year}
\def\<{\thinspace}

\def\boxit#1{\vbox{\hrule\hbox{\vrule\kern3pt\vbox{\kern3pt
	  #1 \kern3pt}\kern3pt\vrule}\hrule}}

\def\BG {Br$\gamma$~}
\def\H2 {H$_2$ 1--0 S(1)~}

\begin{document}
 
\input psfig.sty
\thesaurus{03(08.06.2; 11.09.1; 11.19.1; 11.19.3; 13.09.1)}

\title{Near--infrared line imaging of the circumnuclear starburst rings in 
the active galaxies NGC 1097 and NGC 6574}

\author{J.K. Kotilainen\inst{1}, J. Reunanen\inst{1}, S. Laine\inst{2,3}, 
S.D. Ryder\inst{4,5}} 

\institute{Tuorla Observatory, University of Turku, V\"ais\"al\"antie 20, 
FIN--21500 Piikki\"o, Finland
\and 
Department of Physical Sciences, University of Hertfordshire, College Lane, 
Hatfield, Herts. AL10 9AB, England, U.K.
\and
Present address: Department of Physics and Astronomy, University of Kentucky, 
Lexington, KY 40506--0055, U.S.A
\and
Joint Astronomy Centre, 660 N. A'Ohoku Place, Hilo, HI 96720, U.S.A
\and
Present address: Anglo--Australian Observatory, P.O. Box 296, Epping, NSW 
1710, Australia}

\offprints{J.K. Kotilainen (e--mail: jkotilai@stardust.astro.utu.fi)}

\date{Accepted; received }

\titlerunning{NIR line imaging of NGC 1097 and NGC 6574}
\authorrunning{Kotilainen et al.}

\maketitle

\begin{abstract}

We present high spatial resolution near--infrared broad--band $JHK$ and 
Br$\gamma$ 2.166 $\mu$m and H$_2$ 1--0 S(1) 2.121 $\mu$m emission line images 
of the circumnuclear star formation rings in the LINER/Seyfert 1 galaxy NGC 
1097 and in the Seyfert 2 galaxy NGC 6574. We investigate the morphology, 
extinction, and the star formation properties and history of the rings, by 
comparing the observed properties with an evolutionary population synthesis 
model. 

The clumpy morphology in both galaxies varies strongly with wavelength, due 
to a combination of extinction, emission from hot dust and red supergiants, 
and the age of the stellar populations in the rings. The near--infrared and 
radio morphologies are in general agreement, although there are differences 
in the detailed morphology. From the comparison of Br$\gamma$ and H$\alpha$ 
fluxes, we derive average extinctions toward the hot spots A$_V$ = 1.3 for 
NGC 1097 and A$_V$ = 2.1 for NGC 6574. The observed H$_2$/Br$\gamma$ ratios 
indicate that in both rings the main excitation mechanism of the molecular 
gas is UV radiation from hot young stars, while shocks can contribute only in 
a few regions. 

The starburst rings in both galaxies exhibit small Br$\gamma$ equivalent 
widths. Assuming a constant star formation rate with M$_u$ = 100 M$_\odot$ 
results in extremely long ages (up to 1 Gyr), in disagreement with the 
morphology and the radio spectral index of the galaxies. This situation is 
only slightly remedied by a reduced upper mass cutoff (M$_u$ = 30 M$_\odot$). 
We prefer a model of an instantaneous burst of star formation with M$_u$ = 
100 M$_\odot$ occurring $\sim$6--7 Myr ago. Gaseous nuclear bars parallel to 
the stellar nuclear bar were detected in both galaxies, and we derive M 
$\sim$100 M$_\odot$ for the mass of the excited nuclear H$_2$ emission. 
Finally, we briefly discuss the connection between the rings, bars and the 
fuelling of nuclear activity. 


\end{abstract}

\section{Introduction}

Bright hot spots of star formation (SF) in the circumnuclear regions of 
spiral galaxies are often closely connected to the existence of a bar. In 
many cases, these hot spots make up a kpc--size ring or spiral pattern (e.g. 
Buta \& Crocker 1993). Multiwavelength studies have shown that these 
starburst rings have high SF and supernova rates (e.g. Hummel, van der Hulst 
\& Keel 1987, hereafter H87; Genzel et al. 1995).

The circumnuclear SF rings are thought to arise as a consequence of a 
bar--driven gas inflow and dynamical resonances in the bar (e.g. Combes \& 
Gerin 1985; Friedli \& Martinet 1993; Piner, Stone \& Teuben 1995). Dense gas 
accumulates in shocked regions of orbit crowding along the leading edges of 
the bar, loses angular momentum and flows inward. The gas accumulates in a 
ring between the two inner Lindblad resonances (ILR) and experiences a 
starburst, either through collisions of molecular clouds or by gravitational 
collapse of the ring into dense fragments (e.g. Combes \& Gerin 1985; 
Elmegreen 1994). If no ILRs exist, the gas may continue to flow inward, 
resulting in a nuclear starburst without a ring (Telesco, Dressel \& 
Wolstencroft 1993). 

Although there are plenty of examples of nuclear ring galaxies without 
nuclear activity, suggesting that the nuclear rings and active galactic 
nuclei (AGN) are not necessarily intimately related, there is, however, an 
indicative statistical correlation between the existence of nuclear rings and 
AGN in barred galaxies (Knapen et al. 1999). One of the most important 
questions concerning the nuclear rings is, therefore, that of fuelling of the 
AGN in their centers. Since the large--scale flow slows down substantially at 
the ILR ring, additional mechanisms, such as bar instability within a central 
gas disk ($'$bars within bars$'$; Shlosman, Frank \& Begelman 1989), have 
been suggested to fuel the AGN. 

Multiwavelength studies of AGN with circumnuclear rings can provide insights 
into how the massive star clusters in the rings and nuclear activity are 
related. The relatively unobscured near--infrared (NIR) emission provides a 
better handle than optical emission to quantify the properties of SF in 
galaxies. In this paper we present high resolution broad--band $JHK$ images 
and the first NIR Br$\gamma$ and H$_2$ 1--0 S(1) emission line images of two 
active galaxies with a circumnuclear SF ring, NGC 1097 and NGC 6574. Because 
of their brightness, large angular size and relatively face--on orientation, 
they provide excellent targets for studying these phenomena. Br$\gamma$ 
originates from H II regions surrounding hot young OB star clusters, while 
H$_2$ arises from hot molecular gas and traces the material available for SF. 

NGC 1097 is a nearby (v$_{sys}$ = 1275 km s$^{-1}$; $D=18.2$ Mpc; 1$''$ = 90 
pc; H$_0=$ 70 km s$^{-1}$ Mpc$^{-1}$) SBb galaxy with a strong bar ($PA\simeq 
141^\circ$), inclined at 46$^\circ$ (Storchi--Bergmann, Wilson \& Baldwin 
1996). Non--thermal emission due to shock compression of gas and dust 
associated with the smooth thin dust lanes at the leading edges of the bar is 
clearly visible in radio continuum (Ondrechen \& van der Hulst 1983). NGC 
1097 has an elliptical companion galaxy, NGC 1097A, which influences the 
western spiral arm by tidal interaction (e.g. Ondrechen, van der Hulst \& 
Hummel 1989). NGC 1097 contains a nuclear stellar bar ($PA\simeq 28^\circ$; 
Quillen et al. 1995), almost perpendicular to the primary bar. 
Based on optical emission line ratios, the nucleus of NGC 1097 is usually 
classified as a LINER (Phillips et al. 1984) although the detection of broad, 
double peaked H$\alpha$ emission (Storchi--Bergmann et al. 1993) suggests a 
revised classification as a Seyfert 1. 

The most intriguing feature of NGC 1097 is the bright, almost circular 
circumnuclear SF ring (diameter $\sim$18$''$ = 1.6 kpc), which actually 
consists of two tightly wound clumpy spiral arms. It has been the target of 
extensive multiwavelength studies (e.g. H87; Gerin, Nakai \& Combes 1987: 
Telesco et al. 1993; Barth et al. 1995; Quillen et al. 1995; P\'erez--Olea \& 
Colina 1996; Storchi--Bergmann et al. 1996). According to Storchi--Bergmann 
et al. (1996), the ring of NGC 1097 is situated between the two ILRs 
associated with the primary bar potential, in agreement with numerical 
simulations (Piner et al. 1995). 

NGC 6574 is a little studied SABbc spiral galaxy at v$_{sys}$ = 2282 km 
s$^{-1}$ ($D = 32.6$ Mpc; 1$''$ = 160 pc). It is inclined at $\sim$40$^\circ$ 
and is not strongly interacting with any other major galaxy, as evidenced by 
the symmetric rotation curve (Demoulin \& Chan 1969). Its nucleus is 
classified as a Seyfert 2. Around the nucleus, there is an SF region 
consisting of spiral arms in the form of a wide pseudoring. This ring 
contains $\sim$30 H$\alpha$ emission regions (Gonz\'alez--Delgado et al. 
1997). 

This paper is arranged as follows. In Section 2 we discuss the observations, 
data reduction and the evolutionary model used. In Sections 3 and 4 we 
discuss the morphology of the circumnuclear rings, determine the extinctions 
to the SF complexes from comparison of the recombination line strengths, and 
constrain their SF properties, stellar populations and SF history by a 
detailed comparison among the NIR tracers and with the radio emission. We 
also discuss the origin of the H$_2$ emission in the ring. General discussion 
and conclusions are presented in Section 5. 

\section{Observations, data reduction and the evolutionary model}

\subsection{Observations}

The observations of the Br$\gamma$ 2.1661~$\mu$m and H$_2$ 1--0 S(1) 
2.121~$\mu$m emission lines and the $JHK$ bands were carried out in September 
1998 with the 3.8 m United Kingdom Infrared Telescope (UKIRT) on Mauna Kea, 
Hawaii, under FWHM 0\farcs6--0\farcs7 seeing. We used the 256$\times$256 px 
IRCAM3 camera, with pixel size 0\farcs281 px$^{-1}$ and field of view 
$\sim70''\times70''$. For the emission line observations, we used cooled 
($T=$ 77 K) narrow--band filters and a Fabry--Perot (F--P) etalon with 
spectral resolution $\sim$400 km s$^{-1}$ and equivalent width (EW) 0.0038 
$\mu$m. 

For both emission lines, five 60 sec observations were made at each on--line 
(OL) wavelength and at the nearby blue (BC) and red (RC) continuum, shifted 
with respect to the line centre by $\sim\pm1200$ km s$^{-1}$, following the 
sequence OL--BC--OL--RC and moving the telescope by $\sim$10$''$ in a five 
point grid between the individual integrations for sky subtraction. This 
sequence was repeated until a sufficent on--line integration time was 
achieved. For the $JHK$--bands, five 30 sec integrations were made both 
centered on the galaxy and on sky 5$'$ east. The total integration times for 
the $JHK$--bands and for the \BG and H$_2$ emission lines are 150, 1800 and 
2400 sec (NGC 1097) and 150, 2400 and 2400 sec (NGC 6574). Faint standard 
stars were observed in both broad--band and narrow--band filters immediately 
after the galaxy observations at similar airmass. 

\subsection{Data reduction}

{\sc{IRAF}}\footnote{IRAF is distributed by the National Optical Astronomy 
Observatories, which are operated by the Association of Universities for 
Research in Astronomy, Inc., under cooperative agreement with the National 
Science Foundation.} was used to reduce the observations. The line images 
were divided by a flatfield made from linearized galaxy images, and 
sky--subtracted using the median of the five images at the same wavelength. 
These images were then aligned by using the nuclear position as reference, 
and merged into one on--line image and two continuum images. The alignment of 
the images is accurate to within a fraction of a pixel. Finally, the 
continuum images were scaled with the continuum/line ratio measured from a 
standard star, combined, and subtracted from the on--line images, and the 
final line images were flux calibrated against the spectral type A standard 
stars HD 1160 (M$_K$ = 7.040) and HD 201941 (M$_K$ = 6.626). The reduction of 
the broad--band images consisted of linearization, flatfielding using a sky 
flat, and sky subtraction using the median of five sky images. UKIRT faint 
standard stars FS 29 and FS 35 were used to flux calibrate the broad--band 
images. 

The sensitivity of the F--P is a function of both the line--of--sight 
(l.o.s.) velocity of the galaxy and the shift in the transmitted wavelength 
of the F--P over the field of view. These effects were corrected for by 
dividing the measured fluxes by an inverse Airy function (see e.g. 
Bland--Hawthorn 1995). For NGC 1097, a high resolution l.o.s. velocity field 
exists (Storchi--Bergmann et al. 1996). The only available l.o.s. velocity 
data for NGC 6574 were published by Demoulin \& Chan (1969), who find the 
velocity to be constant to within 20 km s$^{-1}$ at distances 6$''$--18$''$ 
from the nucleus, i.e. encompassing the ring. Therefore, we have assumed that 
the velocity depends only on the angle with respect to the line of nodes and 
the inclination. According to Demoulin \& Chan (1969), the rotational 
amplitude is 240 km s$^{-1}$, inclination 40$^\circ$ and the position angle 
PA = 159$^\circ$. Using these values, we have estimated the velocity field in 
the SF ring. In the nucleus, we assumed the l.o.s. rotational velocity to be 
0 km s$^{-1}$.

The emission linewidths would also need to be considered. This would, 
however, require information about the line profiles in different parts of 
the rings, which is not available. However, usually the spectral lines are 
rather narrow, and this correction is likely to be small. In this paper, the 
emission line {\em images} have not been corrected for by the inverse Airy 
function, because then noise would depend on the position in the image. 
However, the observed emission line {\em fluxes} have been corrected for this 
effect. The observed line emission is weak, so to enhance the S/N, all images 
were smoothed to 1$''$ resolution. 

Photometry in all bands was performed at the location of the detected \BG 
emission regions. The aperture used for each region was selected to include 
as much as possible of the radiation, while avoiding overlap with 
neighbouring regions. The smallest distance between the nearest emission 
regions in these galaxies is $\sim$2$''$, which is larger than the resolution 
of the smoothed images ($\sim$1$''$) and much larger than the seeing during 
the observations ($\sim$0\farcs6). The observed fluxes were corrected for 
Galactic extinction (de Vaucouleurs et al. 1991) and redshift 
(K--correction). We estimate a photometric accuracy in the $JHK$ magnitudes 
of $\sim$0.03 mag, in the $JHK$ colours of $\sim$0.05 mag, and in the 
emission line fluxes of $\sim$10 \%.

\subsection{Extinction}

Optical observations of starburst regions are much more hampered by the large 
and spatially variable extinction than observations in the NIR (A$_K$ 
$\sim$0.1 A$_V$). The derived extinction depends on the assumed geometry of 
the absorbing dust with respect to the emission region. In this paper we 
assume for the sake of simplicity and for easy comparison with previous 
determinations, the case of a foreground screen of dusty cloud filaments. The 
true dust distribution is, however, probably clumpy and mixed with the star 
clusters, resulting in a somewhat larger extinction. Therefore, the derived 
extinction values should be treated as approximations only. 
Note that the exact value of extinction does not affect the EWs of the 
emission lines, if the differential extinction between the lines and the 
continuum is small, as is usually the case in the NIR (Calzetti 1997 and 
references therein). 
Assuming standard interstellar dust properties, case B recombination 
(intrinsic $F_{H\alpha}/F_{Br\gamma}=102$; Osterbrock 1989), and the Landini 
et al. (1984) interstellar extinction curve, $A_\lambda \propto 
\lambda^{-1.85}$, extinction towards the ionized sources can be determined 
from the observed Br$\gamma$ and H$\alpha$ fluxes.

This extinction strictly refers only to the ionized gas. The extinction 
experienced by the continuum can be estimated by comparing the observed 
broad--band colours of the emission regions to the average colours of normal 
unobscured spiral galaxies ($<J--H>$ = 0.75, $<H--K>$ = 0.22; Glass \& 
Moorwood 1985). This method, however, may be biased toward low extinction 
regions, as the most heavily reddened regions may not be detectable in the 
$JHK$ images. Also, the intrinsic colours of the galaxies may be bluer than 
in normal galaxies, leading to an underestimation of the continuum extinction. 

\subsection{The evolutionary Model}

To interpret the SF properties, we have used the stellar population synthesis 
model of Leitherer et al. (1999; hereafter L99), which employs the latest 
stellar evolutionary tracks (e.g. Charbonnel et al. 1999) and stellar 
atmosphere models (e.g. Lejeune, Buser \& Cuisinier 1997). It predicts the 
evolution of a large number of NIR, optical and UV spectral features as a 
function of the burst age, metallicity, and the initial mass function (IMF) 
with lower and upper mass cutoff and slope $\alpha$, for the limiting cases 
of instantaneous burst of SF (ISF) and constant SF rate (CSFR).

The EW of Br$\gamma$, which is sensitive to the age of the starburst, was 
estimated by subtracting a de Vaucouleurs bulge model (depicting the old 
stellar population) from the $K$--band image and dividing the Br$\gamma$ 
fluxes with the remaining $K$--band fluxes. The L99 models predict the number 
of ionizing photons below 912 \AA, N(H$^0$), which can also be estimated from 
the Br$\gamma$ flux as N(H$^0$) $[s^{-1}]$ = 7.63$\times$ 
10$^{13}$L$_{Br\gamma}$[erg s$^{-1}]$. N(H$^0$) allows us to evaluate the 
mass of recently formed stars (in ISF) or the SF rate (SFR) via an assumed 
IMF (in CSFR). N(H$^0$) can also be estimated from the thermal radio emission 
(Condon \& Yin 1990): N(H$^0$) = 7.1$\times$ 10$^{49}$ D$^2$ $\nu^{0.1}$ 
T$_e^{-0.76}$ S$_{th}$, where D is the distance in Mpc, $\nu$ is the 
frequency in GHz, $T_e$ is the electron temperature in 10$^4$ K and $S_{th}$ 
is the thermal radio flux density in mJy. The supernova (SN) rate predicted 
by the L99 models can be compared with the rate derived from non--thermal 
radio emission (Condon \& Yin 1990; based on Galactic SNe): L$_{NT}$ $\sim$ 
1.3$\times$10$^{23}$ $\nu^{-\alpha}$ v$_{SN}$, where $L_{NT}$ is the 
non--thermal radio luminosity in W Hz$^{-1}$, $\nu$ is the frequency in GHz, 
$\alpha$ is the spectral index of non--thermal radiation and v$_{SN}$ is the 
SN rate in yr$^{-1}$. 

Since we do not have enough data for more detailed modelling, and to allow an 
easy comparison with properties derived by previous authors, we assume solar 
metallicity, $\alpha$ = 2.35, and consider two models in what follows: (1) 
ISF with $M_u$ = 100 M$_\odot$ and (2) CSFR with $M_u$ = 30 M$_\odot$. Note 
that our general conclusions remain unchanged if one uses other evolutionary 
models, e.g. those of Cervino \& Mas--Hesse (1994) or Lancon \& 
Rocca--Volmerange (1996). However, the exact age of the starburst strongly 
depends on the chosen burst duration and IMF.

\section{Results}

\subsection{NGC 1097}

\subsubsection{Morphology}

\begin{figure}
\psfig{file=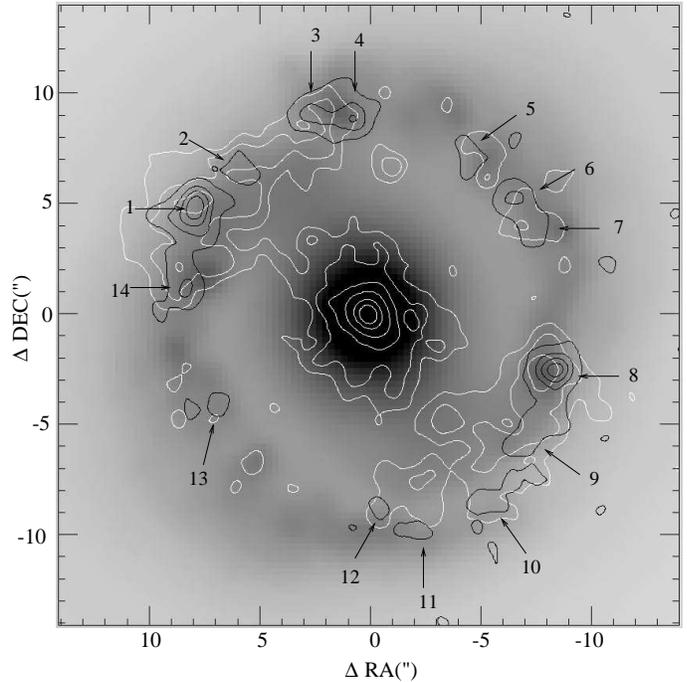,width=9cm,height=9cm}
\caption{The $K$--band image of NGC 1097, overlaid with the \BG emission 
(black contours) and the \H2 emission (white contours). The lowest contour is 
at 16 \% level of the maximum and corresponds to 3$\sigma$. The other 
contours are at 40, 60 and 80 \% (\BG) and at 6.9, 12, 20, 35, 60 and 80 \% 
(H$_2$) of the maximum. The maximum surface brightnesses are $1.1\times 
10^{-15}$ erg s$^{-1}$ cm$^{-2}$ arcsec$^{-2}$ for \BG and $1.7\times 
10^{-15}$ erg s$^{-1}$ cm$^{-2}$ arcsec$^{-2}$ for H$_2$. In this and all 
subsequent figures, north is up and east to the left.}
\end{figure}

\begin{figure}
\psfig{file=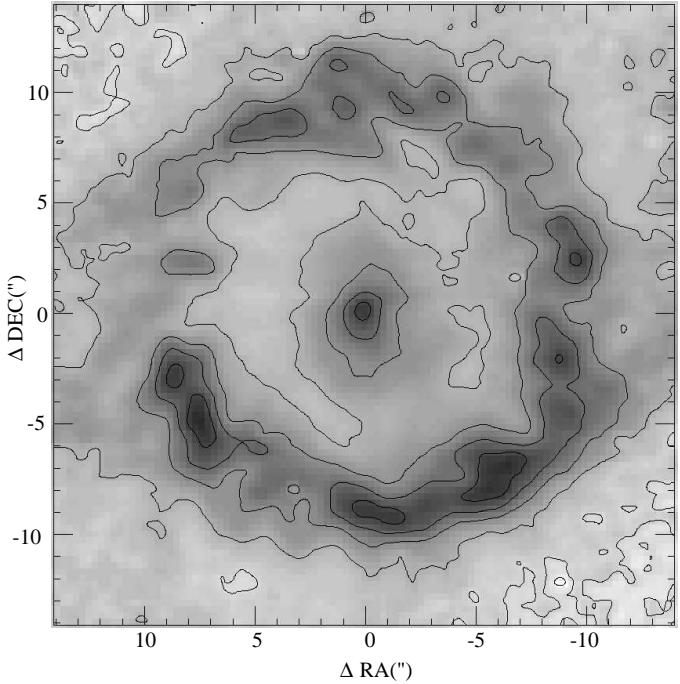,width=9cm,height=9cm}
\caption{The $J$--$K$ colour map of NGC 1097. The highest contour corresponds 
to $J$--$K$ = 1.4 and the other contours are at $J$--$K$ = 0.1 intervals.}
\end{figure}

We have detected 14 emission regions at higher than 3$\sigma$ level in the 
\BG emission line image of NGC 1097 (Fig. 1). There is quite good spatial 
correspondence between the brightest regions in the \BG and radio emission 
(H87; their Fig. 3). Because the non--thermal radio emission is believed to 
arise in SN remnants and the thermal radio emission to be reradiated UV 
emission from OB stars (e.g. Condon 1992), this correspondence is not 
surprising. The detailed differences between radio and \BG can be explained 
as a difference in the ages of the emitting regions, indicating several SF 
generations (see also e.g. NGC 7771, Reunanen et al. 1999). In the $J$--$K$ 
colour map (Fig. 2), the radio peaks correspond quite well to the reddest 
$J$--$K$ colours. The ring is considerably redder than its surrounding 
region. Possible explanations for this are emission from hot dust and red 
supergiants (RSG), or spatially variable extinction. The nucleus is also 
redder than the galaxy. 

There is quite good spatial correlation between the H$\alpha$ (H87; their 
Fig. 5) and the \BG emission, indicating that the effect of extinction is not 
severe in NGC 1097 (see also Section 3.1.2). The worst correspondence is 
toward southeast of the nucleus. Based on radio and H$\alpha$, one would 
expect a \BG emission region between the regions 12 and 13, which is, 
however, not detected. This region has the largest F--P correction factors, 
and it is possible that a \BG emission region exists there, but such a large 
fraction of the emission is shifted outside the F--P filter that it remains 
below our detection limit. 

The correspondence between the H$_2$ and the \BG emission (Fig. 1) is not as 
good as between H$\alpha$ and Br$\gamma$. The clearest difference is in the 
nucleus, which has strong resolved H$_2$ emission, but no detected \BG 
emission (see Section 3.1.3). We derive a nominal upper limit to the nuclear 
\BG emission of $1.5 \times 10^{-16}$ erg s$^{-1}$ cm$^{-2}$ in a 4$''$ 
diameter aperture. In the circumnuclear region, the H$_2$ emission defines a 
broadly similar ring to Br$\gamma$, but there are differences in the detailed 
morphology. 

\begin{figure}
\psfig{file=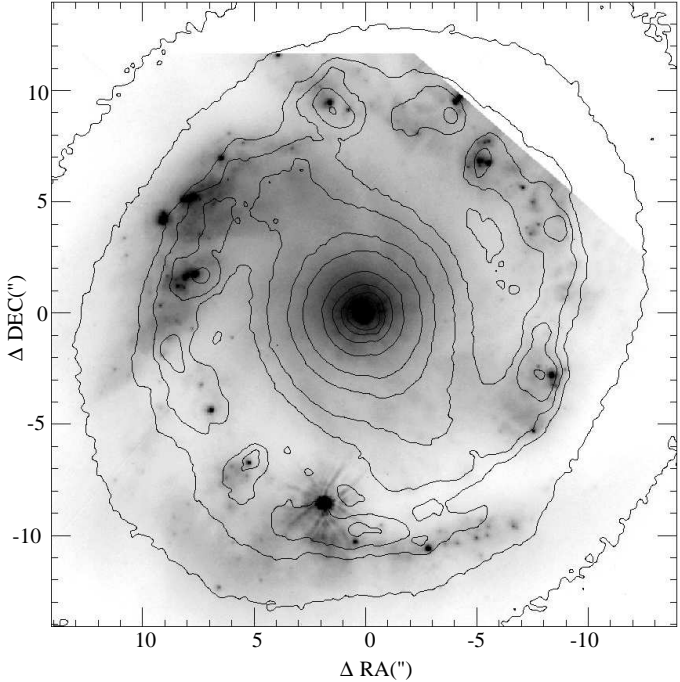,width=9cm,height=9cm}
\caption{The HST $V$--band image of NGC 1097 (Barth et al. 1995), overlaid in 
contours by the 0\farcs65 resolution $K$--band image. The contours are at 
2.5, 5, 10, 12.5, 15, 20, 35, 45, 55, 65 and 80 \% of the $K$--band maximum. 
The strong point source 7$''$ to the south of the nucleus is SN 1992bd.}
\end{figure}

\begin{figure}
\psfig{file=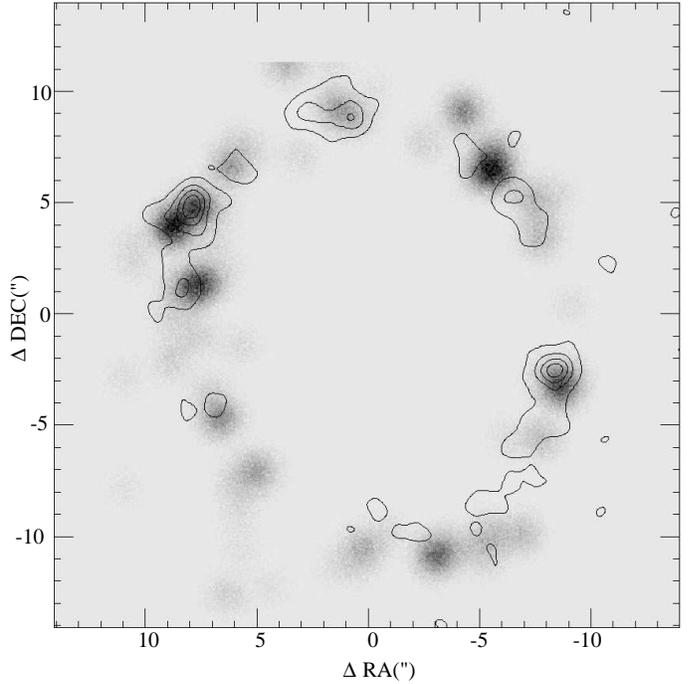,width=9cm,height=9cm}
\caption{The $V$--band HST image from Barth et al. (1995) overlaid in 
contours by the \BG image. The $V$--band image has been smoothed using a 
Gaussian FWHM corresponding to the width of the point source in the $JHK$ and 
\BG images.}
\end{figure}

Barth et al. (1995) observed NGC 1097 with the Hubble Space Telescope (HST) 
and resolved the hot spots in the ring into 88 star clusters. In Fig. 3 we 
show their $V$--band image overlaid in contours by our $K$--band image, and 
in Fig. 4, the \BG emission in contours is overlaid on the $V$--band image, 
which was Gaussian smoothed to the same resolution as \BG. The effective 
radius of the brightest star clusters in the $V$--band, $R_{eff}$ = 3--7 pc, 
indicates that they represent individual SF regions. There is no perfect 
correlation between them and the \BG regions, although most \BG regions have 
a reasonably nearby $V$--band counterpart. There are differences especially 
in the north, where some \BG regions are situated in the $V$--band {\em 
minima}. Neither is there a perfect spatial correspondence between \BG and 
$K$--band (Fig. 1), although some \BG regions (e.g. region 8) have a clear 
counterpart in the $K$--band. In many regions, however, there is overlap 
between the $V$-- and $K$--bands and the \BG. 

The differences between Br$\gamma$, radio, $V$-- and $K$--bands can be 
explained by differences in the age of the emission regions. The $V$--band 
peaks are probably younger than those in the $K$--band, since it takes 3--4 
Myr before the emergence of the first RSGs that dominate the $K$--band 
emission. Less likely, the $K$--band peaks may either be regions of low 
extinction, or young dusty SNe. 

\subsubsection{Star formation properties}

\begin{table*}
\begin{center}
\caption{Br$\gamma$ emission regions in NGC 1097}
\begin{tabular}{lccccccllcccccc}
 & & & \multicolumn{4}{c}{Observed emission}&&\multicolumn{5}{c}{Dereddened emission}\\
\hline
n & ap & corr$^a$ & Br$\gamma$ & H$_2$ & J--H & H--K & A$_V$$^b$ & Br$\gamma$ & H$_2$ & $\frac{H_2}{Br\gamma}$ & J--H & H--K \\
 & $''$ &  & \multicolumn{2}{c}{10$^{-15}$ ergs s$^{-1}$ cm$^{-2}$} & mag & 
mag & mag & \multicolumn{2}{c}{10$^{-15}$ ergs s$^{-1}$ cm$^{-2}$} & & mag & 
mag\\
1 & 3.4& 1.12& 4.43& 2.68& 0.69& 0.28& 1.0& 4.76& 2.88& 0.61& 0.58& 0.23\\
2 & 2.0& 1.01& 0.68& 0.72& 0.70& 0.26& 1.1& 0.74& 0.78& 1.05& 0.58& 0.20\\
3 & 1.7& 1.62& 1.59& 0.67& 0.74& 0.34& 1.3& 1.75& 0.74& 0.42& 0.60& 0.27\\
4 & 1.7& 1.69& 1.87& 0.46& 0.72& 0.37& 1.3& 2.06& 0.50& 0.25& 0.59& 0.30\\
5 & 2.0& 1.71& 0.90& 0.51& 0.74& 0.30& 1.2& 0.98& 0.56& 0.58& 0.61& 0.23\\
6 & 1.4& 1.75& 1.05& 0.34& 0.72& 0.30& 1.5& 1.31& 0.37& 0.32& 0.57& 0.22\\
7 & 1.4& 1.81& 0.78& 0.51& 0.74& 0.31& 1.5& 0.87& 0.57& 0.63& 0.58& 0.23\\
8 & 3.4& 1.17& 2.96& 2.68& 0.79& 0.32& 1.8& 3.39& 3.06& 0.91& 0.61& 0.22\\
9 & 3.4& 1.06& 1.09& 1.84& 0.78& 0.31& 1.5& 1.22& 2.06& 1.70& 0.62& 0.23\\
10& 3.4& 1.06& 0.85& 1.16& 0.79& 0.33& 1.2& 0.93& 1.27& 1.36& 0.66& 0.27\\
11& 2.0& 2.04& 0.81& 0.43& 0.78& 0.33& 0.1& 0.82& 0.43& 0.52& 0.74& 0.32\\
12& 2.0& 1.98& 0.71& 0.73& 0.81& 0.37& 0.3& 0.73& 0.75& 1.04& 0.76& 0.35\\
13& 3.4& 3.29& 2.86& 1.01& 0.76& 0.34& 2.2& 3.37& 1.20& 0.36& 0.55& 0.21\\
14& 3.4& 1.79& 3.57& 2.74& 0.70& 0.27& 2.2& 4.21& 3.24& 0.77& 0.48& 0.15\\
nucl & 1.7& 1.07&$<$0.15&2.68&0.84& 0.34&    &      &      &$>$17.9& &     \\
\hline
\end{tabular}
\end{center}
$^a$: Correction factor for the velocity field and the change of wavelength 
across the array.\\
$^b$: Determined from the H$\alpha$/Br$\gamma$ ratio, assuming $A_\lambda 
\propto \lambda^{-1.85}$ (Landini et al. 1984).\\
\end{table*}
\normalsize

In Table 1 we give for the 14 \BG emission regions the aperture diameter in 
arcsec, the calibration coefficient from the Airy function, the \BG and H$_2$ 
fluxes, and the $JHK$ colours. The smallest values of FWHM correspond to a 
size of the emitting region of 0\farcs8 (70 pc). Therefore, the emission 
regions detected in the NIR actually are conglomerates of several OB 
associations and giant molecular clouds (c.f. $R_{eff}=3--7$ pc; Barth et al. 
1995), probably similar to scaled--up versions of the 30 Dor H II region in 
the LMC (e.g. Walborn et al. 1999). 

Evans et al. (1996) gave for the H$\alpha$ emission regions in NGC 1097 the 
fluxes in instrumental units, but did not report the transformation 
coefficient into physical units. We have, therefore, calibrated the Evans et 
al. image using the lower resolution H$\alpha$ image (H87) which, however, 
partly includes upper limits. Following the method outlined in Section 2.3., 
we derive for the emission regions in NGC 1097 extinctions $A_V$ = 0.1--2.2 
(average $A_V$ = 1.3). H87 derived, based on H$\alpha/$H$\beta$, for the 
eastern ($\sim$ region 1) and western ($\sim$ region 8) parts of the ring 
extinctions $A_V=0.85$ and $A_V=2.83$, respectively. Other extinction 
determinations for NGC 1097 include average A$_V$ = 1.1 for the ring (Barth 
et al. 1995), and A$_V$ = 0.6--3.0 for various parts of the ring (Walsh et 
al. 1986). Our extinction determination is in good agreement with previous 
results, while the large spread between the emission regions indicates large 
spatial variation in the extinction within the ring. The 
extinction--corrected fluxes of the \BG emission regions are given in Table 1.

The H$_2$/Br$\gamma$ ratio gives important clues about the excitation 
mechanism(s) of the hot molecular gas (Puxley, Hawarden \& Mountain 1990). 
The main mechanisms suggested are thermal excitation in hot gas by low 
velocity shocks (e.g. Draine, Roberge \& Dalgarno 1983) or by intense X--ray 
radiation (e.g. Maloney, Hollenbach \& Tielens 1996), and fluorescent 
excitation by strong UV radiation (e.g. Black \& van Dishoeck 1987). The 
observed H$_2$/Br$\gamma$ ratios for NGC 1097 (0.25--1.70, average 0.75) can 
mainly be explained with UV excitation by young stars (H$_2$/Br$\gamma\simeq 
0.4--0.9$; Puxley et al. 1990). In the regions where the ratio is larger than 
0.9, the density of the gas clouds may be greater, the upper mass cutoff 
higher, or there may be a contribution from shock excitation, possibly due to 
supernovae. In fact, the regions with the largest H$_2$/Br$\gamma$ ratio 
(regions 2 and 9), are located at the opposite ends of the bar. This agrees 
well with the idea of a propagating starburst (e.g. Rieke et al. 1993), where 
SF begins in the nucleus and propagates into the ends of the bar and into the 
circumnuclear ring through shocks in the bar. Further data, e.g. [FeII] 1.644 
$\mu$m, are required to confirm the excitation mechanism(s). 

\begin{table*}
\begin{center}
\caption{Star formation properties of NGC 1097}
\begin{tabular}{lcccccccccr}
 & & & \multicolumn{4}{c}{instantaneous star formation}  & \multicolumn{4}{c}{constant star formation rate}\\
\hline
n & N(H$^0$) & EW & age & mass & SFR$^a$ & v$_{SN}$ & age & mass$^b$ & SFR & v$_{SN}$ \\
 & 10$^{52}$ s$^{-1}$ & \AA & Myr & 10$^{6}$ M$_\odot$ & M$_\odot$ yr$^{-1}$ 
& 10$^{-3}$ yr$^{-1}$ & Myr & 10$^{6}$ M$_\odot$ & M$_\odot$ yr$^{-1}$ & 
10$^{-3}$ yr$^{-1}$\\ 
 1& 2.82& 30.3& 6.23& ~6.56& 0.95& ~6.4& ~9.56& 3.0 & 0.31& 1.0 \\
 2& 0.44& ~8.5& 6.63& ~1.74& 0.26& ~1.6& 49.2& 2.3 & 0.05& 1.0 \\ 
 3& 1.04& 19.8& 6.36& ~3.06& 0.48& ~2.9& 11.4& 1.3 & 0.11& 0.6 \\
 4& 1.22& 20.2& 6.35& ~3.53& 0.56& ~3.3& 11.3& 1.5 & 0.13& 0.6 \\
 5& 0.58& 11.7& 6.53& ~2.08& 0.32& ~1.8& 24.9& 1.6 & 0.07& 0.9 \\
 6& 0.78& 18.9& 6.37& ~2.33& 0.37& ~2.2& 11.7& 1.0 & 0.09& 0.5 \\
 7& 0.52& 12.3& 6.51& ~1.82& 0.28& ~1.6& 22.6& 1.3 & 0.06& 0.7 \\
 8& 2.01& 16.8& 6.40& ~6.39& 1.00& ~6.0& 12.8& 2.8 & 0.22& 1.3 \\
 9& 0.72& ~8.1& 6.65& ~2.99& 0.45& ~2.8& 56.2& 4.4 & 0.08& 1.5 \\
10& 0.55& ~8.1& 6.65& ~2.29& 0.34& ~2.1& 56.2& 3.4 & 0.06& 1.1 \\
11& 0.49& 10.6& 6.57& ~1.79& 0.27& ~1.6& 30.0& 1.6 & 0.05& 0.8 \\
12& 0.43& 12.1& 6.52& ~1.72& 0.26& ~1.5& 23.3& 1.1 & 0.05& 0.6 \\
13& 2.00& 19.8& 6.36& ~5.89& 0.93& ~5.6& 11.4& 2.5 & 0.22& 1.1 \\
14& 2.50& 14.3& 6.46& ~8.51& 1.32& ~7.8& 16.0& 4.4 & 0.28& 2.2 \\
total&   &      &      & 51.0 & 7.8 & 47.0 &      & 32.2 & 1.8 & 10.0 \\
\hline
\end{tabular}
\end{center}
$^a$: Mass divided by age\\
$^b$: SFR multiplied by age\\
\end{table*}
\normalsize

We have determined the SF properties of NGC 1097 by comparing the observed 
quantities with the L99 models (Table 2). The \BG EWs are small (8.1--30 \AA, 
average 15 \AA). Assuming the CSFR model with M$_u=100$ M$_\odot$, we derive 
very long ages for the burst, up to 1 Gyr, which are inconsistent with the 
clumpy $K$--band morphology of NGC 1097. During such a long time, these 
irregularities would have had time to smooth out. Also, the lifetime of 
individual giant molecular clouds in our Galaxy is only $\sim$20--40 Myr, 
after which turbulence and heating from SNe disrupt them and inhibit further 
SF (Blitz 1991). Finally, gas consumption during such a long time with high 
mass SF would by far exceed the theoretical mass inflow rate into a 
circumnuclear ring (Piner et al. 1995). On the other hand, assuming a reduced 
upper limit to the mass function, M$_u=30$ M$_\odot$, we derive a much lower 
duration, 9.6--56 Myr, for the SF. The corresponding SFR is 0.05--0.31 
M$_\odot$ yr$^{-1}$ (average 0.13 M$_\odot$ yr$^{-1}$, total within the ring 
1.8 M$_\odot$ yr$^{-1}$) and SN rate is 0.5--2.2 $\times 10^{-3}$ yr$^{-1}$ 
(average 1.0$\times 10^{-3}$ yr$^{-1}$, total within the ring 1.4 $\times 
10^{-2}$ yr$^{-1}$). 

Assuming the ISF model, we derive even shorter ages, 6.2--6.7 Myr. The 
corresponding masses of the emission regions are 1.7--8.5 $\times 10^6$ 
M$_\odot$ (average 3.6 $\times 10^6$ M$_\odot$, total within the ring 5.1 
$\times 10^7$ M$_\odot$) and the SN rate is $V_{SN}$ = 1.5--7.8 $\times 
10^{-3}$ yr$^{-1}$ (average 3.4 $\times 10^{-3}$ yr$^{-1}$, total 4.7 $\times 
10^{-2}$ yr$^{-1}$). The total mass of the hot gas, although large, is still 
only a small fraction of both the total stellar mass in the ring, $\sim$1 
$\times$ 10$^9$ M$_\odot$ (Quillen et al. 1995) and the total molecular gas 
mass, M(H$_2$) = 1.3 $\times$ 10$^9$ M$_\odot$ (Gerin et al. 1987). 

\begin{figure}
\psfig{file=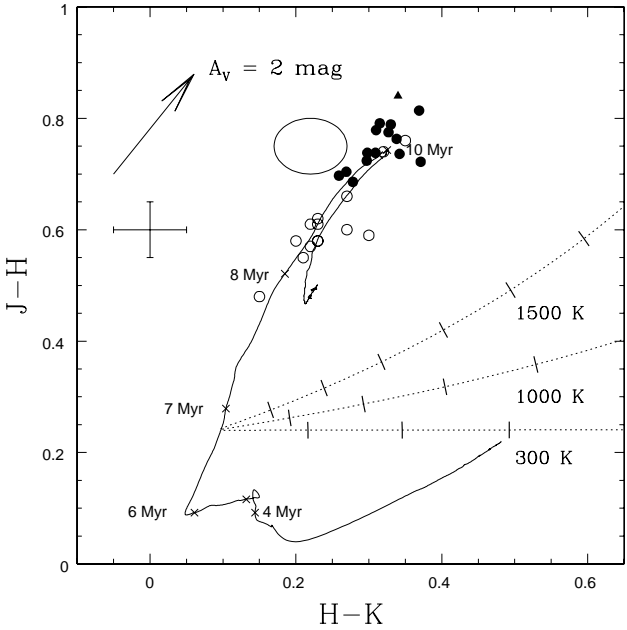,width=9cm,height=9cm}
\caption{The observed (filled circles) and extinction--corrected (open 
circles) $J$--$H$ and $H$--$K$ colours of NGC 1097. The nucleus is marked as 
a triangle. The ellipse shows the average colours of unobscured spiral 
galaxies (Glass \& Moorwood 1985). The arrow shows the effect of extinction 
of A$_V=$ 2 mag, and below the arrow are the photometric errors of the 
colours. The continuous line shows the L99 model colours for an instantaneous 
burst of star formation ($\alpha$=2.35, M$_u$ = 100 M$_\odot$). The dashed 
lines show the effect of hot dust emission on the colours, with the tickmarks 
marking the fractional contribution of dust to the $K$--band emission at 10 
\% intervals.}
\end{figure}

The reliability of conclusions based on $JHK$ colours is reduced by our 
incomplete knowledge of the evolution and properties of RSGs (Origlia et al. 
1999). The hot spots in the SF ring of NGC 1097 are much redder than their 
environment, although young stars ($t\sim 6$ Myr) should be blue. Only when 
the RSGs appear ($t\sim10$ Myr), the colours become redder. Fig. 5 shows the 
$J$--$H$ and $H$--$K$ colour diagram of the \BG emission regions in NGC 1097, 
including the average $J$--$H$ and $H$--$K$ colours of normal unobscured 
spiral galaxies, and the effect of emission from hot dust at various 
temperatures. The extinction--corrected colours agree rather well with the 
L99 evolutionary model. The differences of the $H$--$K$ colour with respect 
to the L99 model can be explained by assuming on average $\sim$20 \% 
fractional contribution to the $K$--band emission from moderately hot dust. 
The $J$--$H$ colour resembles the colour of a 8--9 Myr, not $\sim$6 Myr, old 
ISF model. These ages, however, agree quite well, because the ISF model gives 
in fact a lower limit to the age of the burst, as the burst is not 
instantaneous, but it has a short, finite duration (shorter than the lifetime 
of the forming stars; $t$ $<$ a few 10$^6$ yr). Furthermore, the NIR colours 
may be contaminated by older starburst generations.

\begin{table*}
\begin{center}
\caption{Radio emission of NGC 1097 (from Hummel et al. 1987)}
\begin{tabular}{lccccccc}
\hline
n$^a$ & n$^b$ & S$_{4.885 GHz}$ & $\alpha$ & S$_{th}$/S$_{tot}$ & S$_{th}$ & v$_{SN}$ & N(H$^0$)$^c$ \\
 & & mJy & & & mJy & 10$^{-3}$ yr$^{-1}$ & 10$^{52}$ \\ 
  1     & 3 4 & 5.36 & -0.46 & 0.56 & 3.0 & 2.4 & 8.3\\
  2$^d$ & 2 & 2.66 & -0.58 & 0.37 & 1.1 & 1.7 & 2.7  \\
  3     & 1 & 3.34 & -0.58 & 0.37 & 1.1 & 2.2 & 3.4  \\
  4$^d$ & 14& 1.76 & -0.97 & $-$  & $-$ & 1.8 & --   \\
  5     & 13& 5.36 & -0.54 & 0.44 & 2.3 & 3.1 & 6.5\\
  6     & --& 5.44 & -0.62 & 0.30 & 1.5 & 3.9 & 4.6  \\
  7     & 12& 3.86 & -0.67 & 0.21 & 0.8 & 3.1 & 4.6  \\
  8     & 8 & 5.74 & -0.53 & 0.45 & 2.6 & 3.3 & 7.1\\
  9$^d$ & -- & 2.66 & -0.76 & 0.02 & $-$ & 2.8 & --  \\
 10     & 6 7 & 4.31 & -0.51 & 0.48 & 2.3 & 2.3 & 5.7\\
total        &  &      &       &      &     & 26.6& 43 \\
\hline
\end{tabular}
\end{center}
$^a$: Labeling by Hummel et al. (1987)\\
$^b$: The nearest \BG emission region\\
$^c$: Determined assuming $T_e = 10^4 K$\\
$^d$: Between the 1.465 GHz radio emission regions\\
\end{table*}
\normalsize

Table 3 gives information about the radio emission in NGC 1097 (from H87), 
from which we have estimated the number of ionizing photons N(H$^0$) and the 
SN rate $v_{SN}$. Most radio emission regions in NGC 1097 are dominated by 
non--thermal emission (on average 2/3 of the total). The total SN rate 
derived from the non--thermal radio emission is 0.053 yr$^{-1}$, which is in 
reasonable agreement with the value derived from Br$\gamma$, 0.099 yr$^{-1}$. 
The latter value depends, however, on the adopted IMF parameters $\alpha$ and 
$M_u$. The N(H$^0$) derived from the \BG emission is $\simeq 23\times 
10^{52}$ s$^{-1}$, whereas N(H$^0$) derived from the thermal 4.885 GHz radio 
emission is $\simeq 43\times 10^{52}$ s$^{-1}$. Note that an even larger 
difference between the N(H$^0$) derived from radio and \BG was found for the 
starburst galaxy with a circumnuclear ring, NGC 7771 (Reunanen et al. 1999). 
For NGC 7771, we suggested that the fraction of the thermal radio emission 
had been overestimated, and the same may be applicable to NGC 1097.  

The radio spectral index of NGC 1097 is between --0.46 and --0.97 within the 
ring (H87). According to the model of Mas--Hesse \& Kunth (1991), the 
spectral indices observed for NGC 1097 are typical for an ISF of 5--9 Myr 
age. For the CSFR model, the steepest possible spectral index is 
$\sim$--0.25. This strongly supports the conclusion that the ISF model is 
more likely to be applicable to NGC 1097. Further support for this comes from 
the clumpy $K$--band morphology (see above).

\subsubsection{Nuclear emission}

No \BG emission was detected in the nucleus of NGC 1097, suggesting that it 
does not contain large amounts of massive young stars. The nucleus, however, 
is a strong source of H$_2$ emission (L(H$_2$) = 1.3 $\times$ 10$^5$ 
L$_\odot$ in 10$''$ aperture), implying abundant fuel for SF. This is 
supported by the nuclear $JHK$ colours, which are much redder than the 
colours of the bulge and resemble the colours of the ring. The simplest 
explanation for these colours is the presence of hot dust coexisting with the 
dense molecular gas. 

The nuclear source of H$_2$ emission in NGC 1097 (Fig. 1) has a FWHM 
0\farcs94 (85 pc). Comparing this with the Gaussian FWHM resolution of the 
image (0\farcs65), we derive for the true extent of the resolved nuclear 
source $\sim$0\farcs7 (65 pc). The PA of the resolved H$_2$ emission is 
$\sim$50$^\circ$, whereas the PA of the stellar $K$--band bar is $\simeq 
28^\circ$. Therefore, the H$_2$ emission is nearly elongated along but 
leading the stellar bar. Although kinematical data are required to verify its 
reality, the H$_2$ emission probably is in the form of a gaseous nuclear bar, 
and it appears that gas is flowing into the nucleus along the nuclear bar. 
Since the ellipticity of the isophotes is much larger in the H$_2$ than in 
the $K$--band, the gaseous H$_2$ bar is narrower than the stellar bar. 

The average surface brightness of the nuclear H$_2$ emission in a 4$''$ (360 
pc) aperture is 2.1 $\times$ 10$^{-8}$ W cm$^{-2}$ sr$^{-1}$. Following the 
method of Meaburn et al. (1998), we derive $\sim$120 M$_\odot$ for the mass 
of the excited H$_2$. This value should be multiplied by an unknown but 
probably small factor for the linewidth, which may be broader than the width 
of the F--P passband (see Section 2.2). The resulting mass, M $\sim$120 
M$_\odot$, is similar to those derived for the Seyfert galaxies NGC 3079 (800 
M$_\odot$; Meaburn et al. 1998) and NGC 3227 (250 M$_\odot$; Fernandez et al. 
1999). In all cases, the mass of the excited H$_2$ is only a small fraction 
of the available molecular gas mass.

Because the smallest \BG EWs (oldest ages) are found near the ends of the 
nuclear bar, SF may have commenced in the nucleus (where it has already 
ended) and propagated into the ring through the nuclear bar. This possibility 
is also supported by the H$_2$/Br$\gamma$ ratios in the ring, which indicate 
that shocks dominate the excitation of H$_2$ near the ends of the nuclear 
bar (Section 3.1.2). Similarly, the SN rate/SFR ratio is largest in the hot 
spots at the ends of the nuclear bar, suggesting that SF there is decaying 
(i.e. fewer massive stars are produced, while the SN rate continues 
unchanged). Against the idea of a propagating starburst, we note that there 
is no strong evidence for recent star formation in the region between the 
nucleus and the ring either in Br$\gamma$ (this work) or in radio and 
H$\alpha$ emission (H87). However, as studies of collisional ring galaxies 
(e.g. Appleton \& Struck-Marcell 1996) and HI supershells (e.g. Ryder et al. 
1995 and references therein) have shown, evidence of triggered bursts of 
massive star formation (e.g. X-rays or [S II] emission from SN remnants) can 
fade quickly.

Alternatively, the nuclear gaseous/stellar bar found in this study may 
be driving gas toward the nucleus to fuel the Seyfert. This is suggested 
by the models of e.g. Pfenniger \& Norman (1990) and Friedli \& 
Martinet (1993), in which the nuclear bar decouples from the primary bar 
and has a higher pattern speed. 
Such a short--lived ($\sim$10$^8$ yr) nuclear bar would shock the 
circumnuclear gas at its leading edges, resulting in further gas inflow. The 
gaseous nuclear bar in NGC 1097 (and in NGC 6574, see Section 3.2.1) adds to 
the increasing number of such cases known in literature (e.g. NGC 3351, 
Devereux, Kenney \& Young 1992; NGC 1808, Kotilainen et al. 1996; NGC 1068, 
Davies et al. 1998). 

\subsection{NGC 6574}

\subsubsection{Morphology}

\begin{figure}
\psfig{file=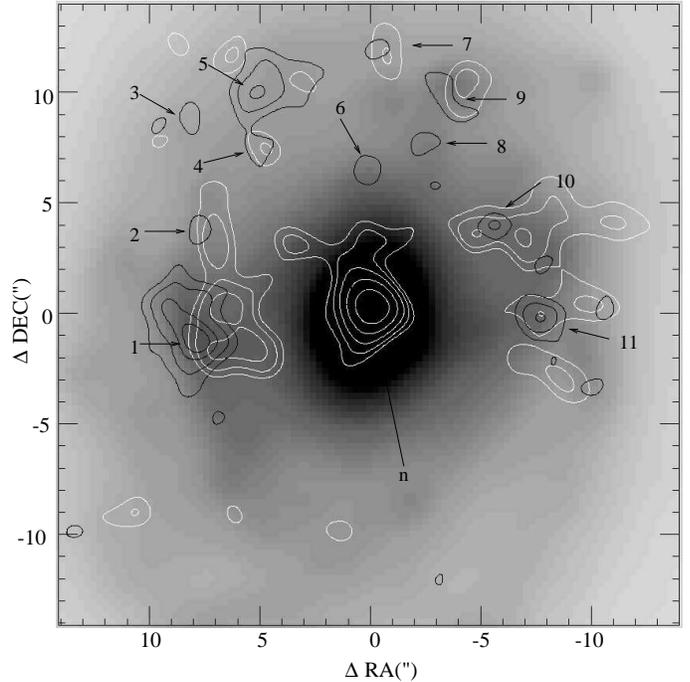,width=9cm,height=9cm}
\caption{The $K$--band image of NGC 6574 overlaid with \BG (black contours) 
and \H2 (white contours) emission. The \BG and H$_2$ contours are at 31.5, 
50, 70 and 85 \%, and at 27.6, 35, 45, 60 and 85 \% from the maximum level, 
respectively. The lowest contours are at 3$\sigma$ level. The maximum surface 
brightnesses are 2.9$\times 10^{-16}$ erg s$^{-1}$ cm$^{-2}$ arcsec$^{-2}$ 
for \BG and 1.6$\times 10^{-16}$ erg s$^{-1}$ cm$^{-2}$ arcsec$^{-2}$ for 
H$_2$.}
\end{figure}

\begin{figure}
\psfig{file=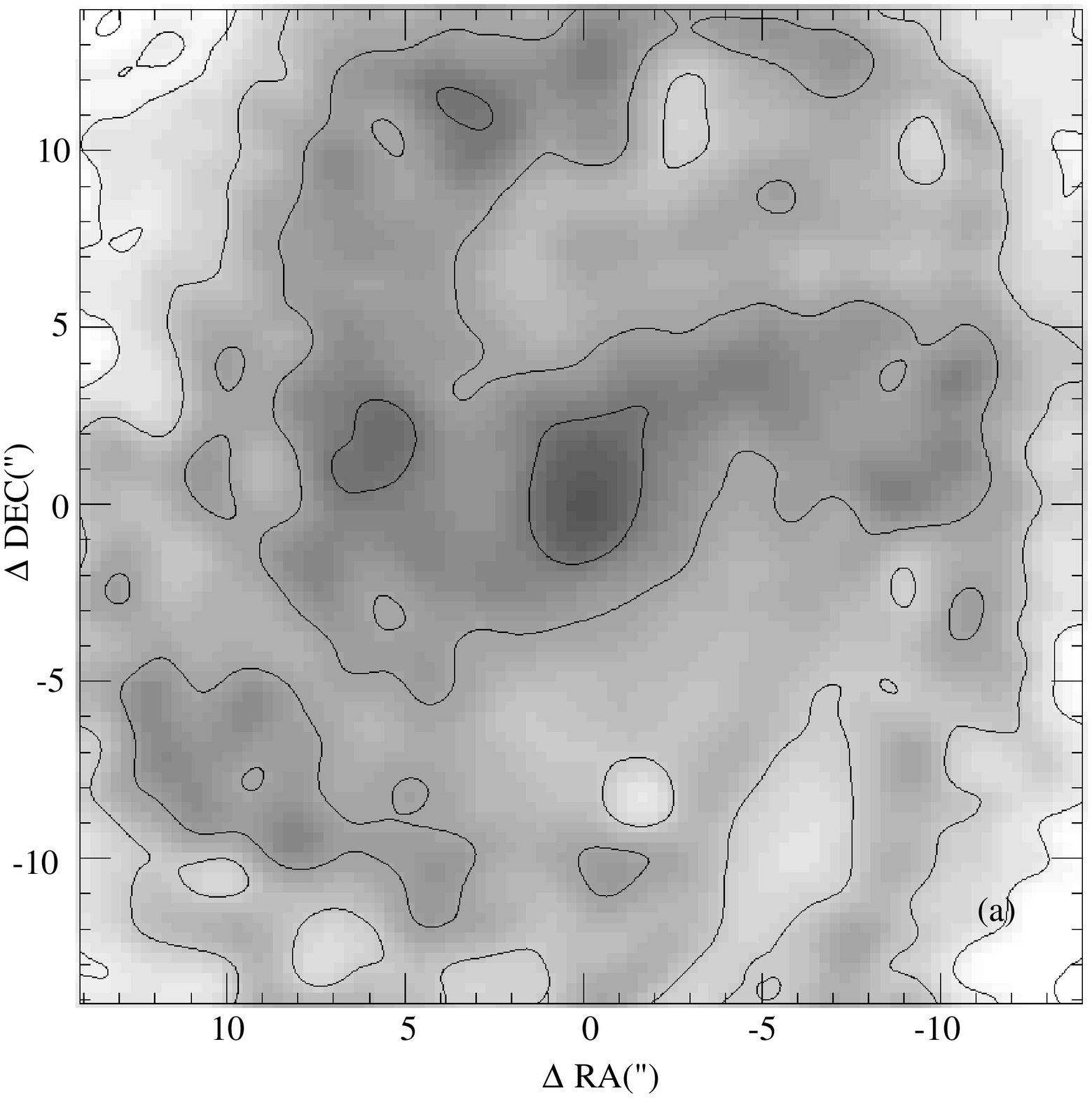,width=9cm,height=9cm}
\psfig{file=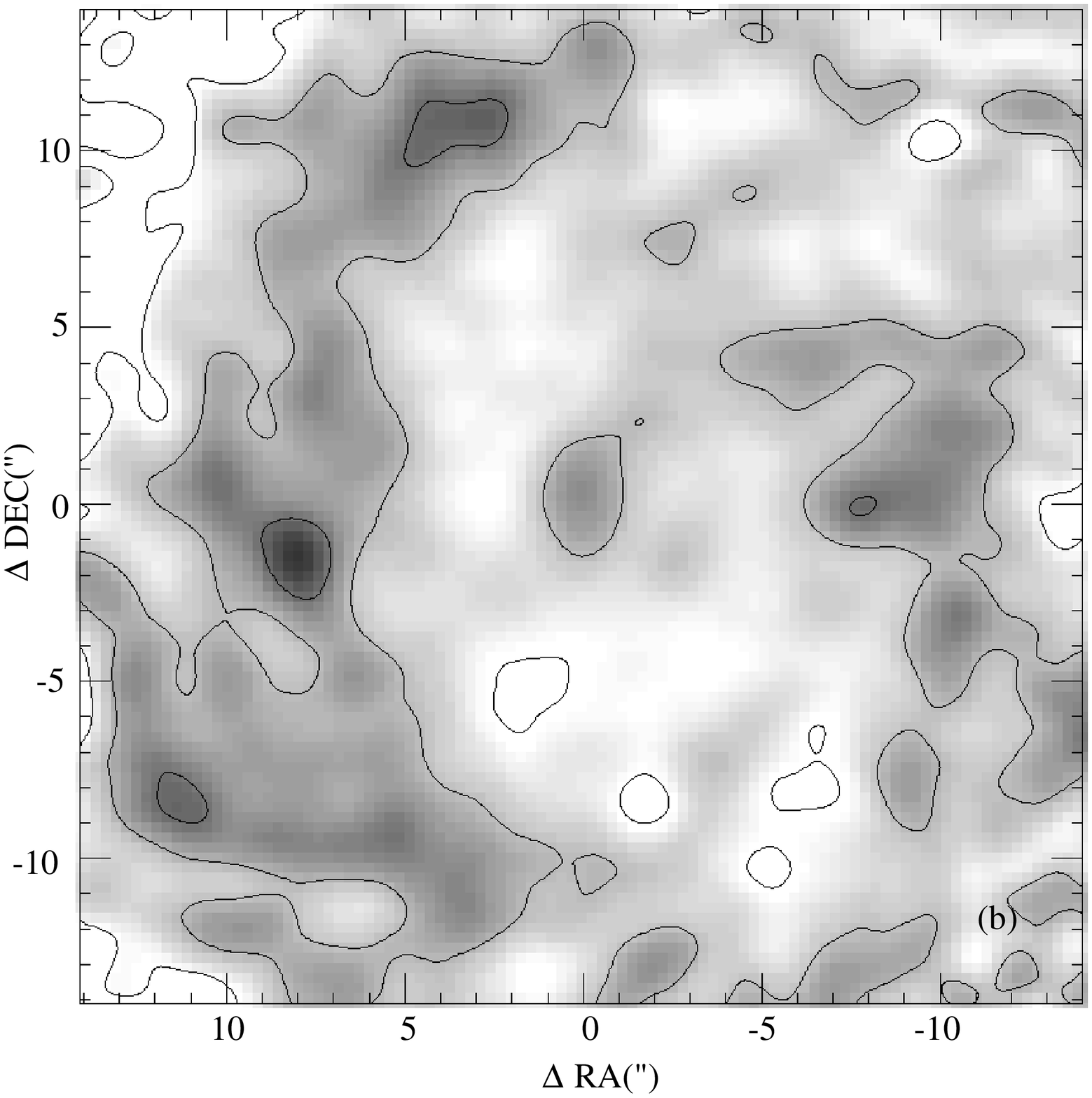,width=9cm,height=9cm}
\caption{The {\bf a)} $J$--$H$ and {\bf b)} $H$--$K$ colour maps of NGC 6574. 
The highest contours correspond to $J$--$H$ = 0.8 and $H$--$K$ = 0.35. The 
other contours are at $J$--$H$ = $H$--$K$ = 0.05 intervals.}
\end{figure}

\begin{figure}
\psfig{file=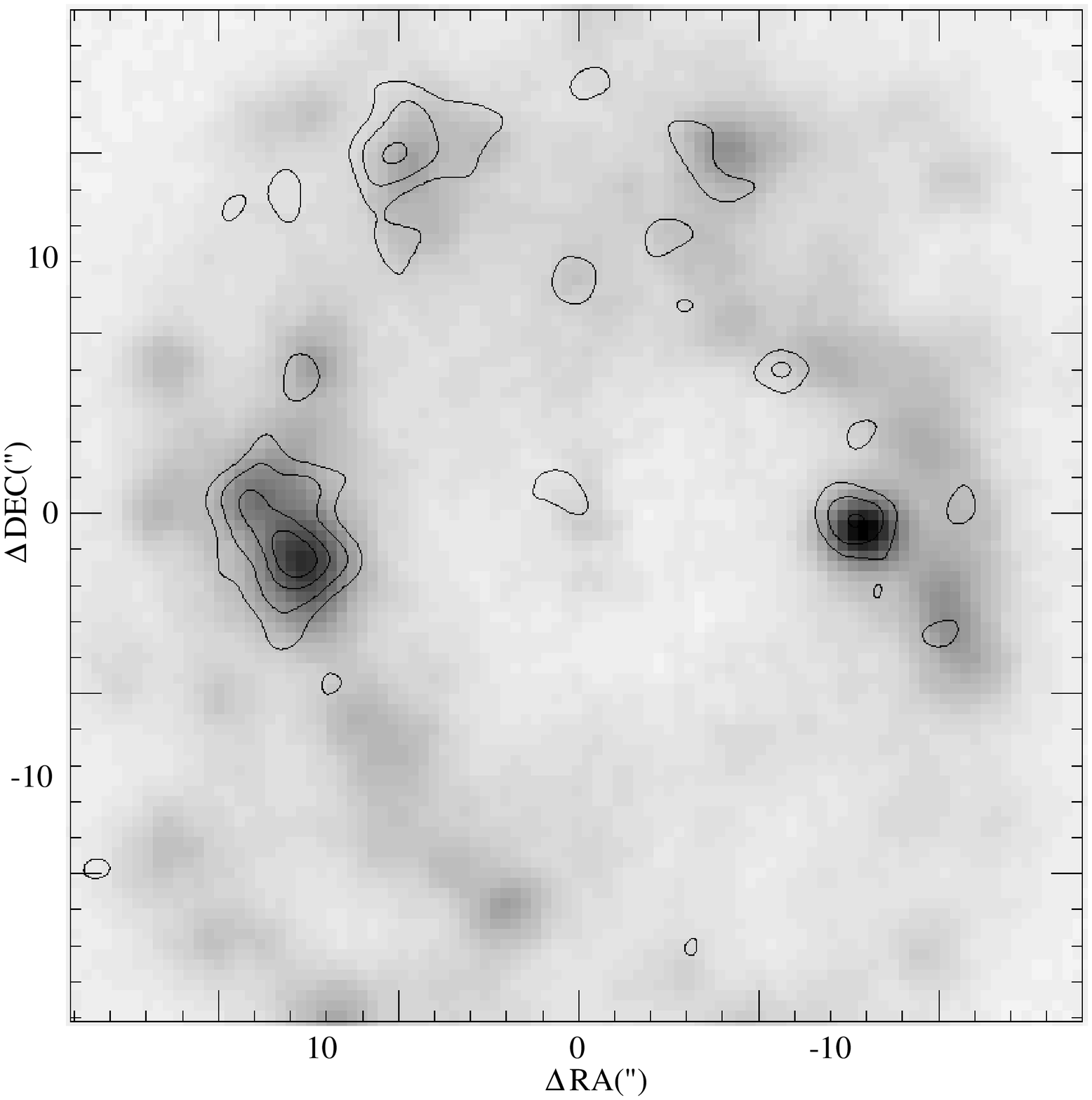,width=9cm,height=9cm}
\caption{The H$\alpha$ image of NGC 6574 (from Gonz\'alez--Delgado et al. 
1997), overlaid in contours by the \BG image.}
\end{figure}

The $K$--band, \BG and H$_2$ images of the nuclear region of NGC 6574 are 
shown in Fig. 6. We have detected 12 regions (including the nucleus) at 
higher than 3$\sigma$ level in the \BG emission. The spiral arms of NGC 6574 
are clearly visible in the $K$--band, while the outer bar 
(PA $=$ 105$^\circ$) can be seen in the $J$--$H$ colour map, but not in the 
$H$--$K$ map (Fig. 7). Fig. 8 shows the H$\alpha$ image (from 
Gonz\'alez--Delgado et al. 1997), overlaid in contours with the \BG image. 
Both the \BG and H$\alpha$ emission regions are located in the spiral arms. 
Due to the higher S/N, H$\alpha$ emission regions are visible all along the 
spiral arms, whereas only the brightest \BG regions can be seen. Many 
emission line regions have a counterpart in the $K$--band (Fig. 6), but the 
spatial correspondence is not perfect. 

The morphology of the H$_2$ emission is different from that of \BG (Fig. 6). 
East of the nucleus, near the end of the bar, there is the strongest region 
in both H$_2$ and Br$\gamma$, but the H$_2$ peak is closer to the nucleus. In 
northeast, \BG is strong whereas H$_2$ is weak and diffuse. In the west, both 
emission lines are situated in the same regions, but H$_2$ is more broadly 
distributed. To the south of the nucleus there is neither H$_2$ nor 
Br$\gamma$ emission, but this may be an artifact due to the narrow bandwidth 
of the F--P. The H$_2$ emission is elongated at the ends of the bar 
perpendicular to the bar. The nucleus is a strong source of H$_2$ emission 
(L(H$_2$) = 4.6 $\times$ 10$^4$ L$_\odot$ in 10$''$ aperture), but there is 
only weak \BG emission. The FWHM of the nuclear H$_2$ source is 
$\sim$1\farcs8 (290 pc), much larger than the seeing FWHM $\sim$0\farcs7. 
This resolved nuclear H$_2$ emission (Fig. 6) is closely parallel to but 
leading the nuclear stellar bar visible in the $K$--band (PA $\simeq$ 
150$^\circ$), and probably forms a gaseous nuclear bar. To the northwest and 
west of the nucleus there is an arc--like region of H$_2$ emission. 

The average surface brightness of the nuclear H$_2$ emission in a 5$''$ (800 
pc) aperture is 2.6 $\times$ 10$^{-9}$ W cm$^{-2}$ sr$^{-1}$. Following the 
method described in Section 3.1.2, we derive $\sim$80 M$_\odot$ for the mass 
of the excited H$_2$, similar to that derived for NGC 1097 above.

\begin{figure}
\psfig{file=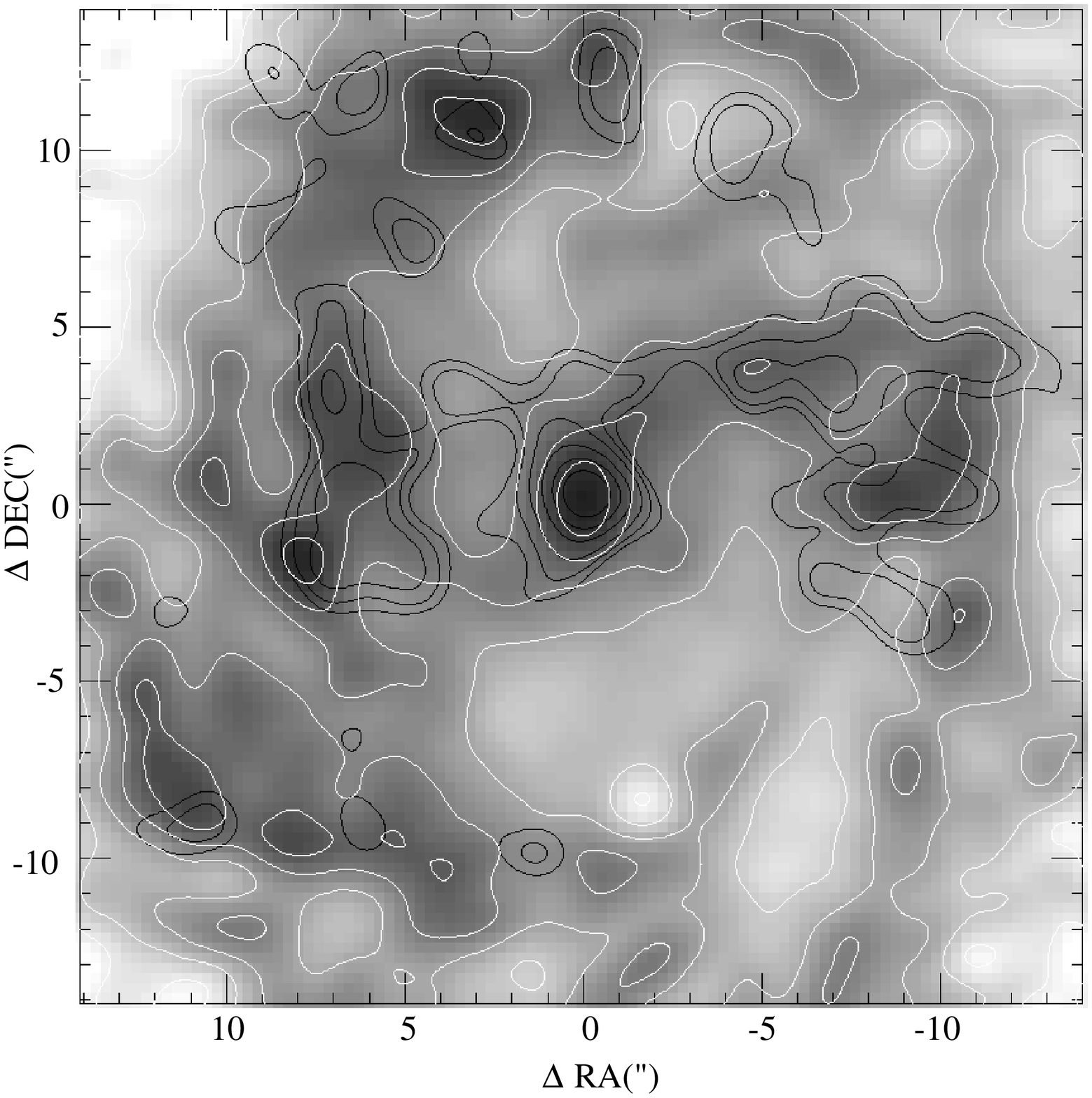,width=9cm,height=9cm}
\caption{The $J$--$K$ colour map of NGC 6574 highlighted in white contours. 
Overlaid in black contours is the \H2 image, with the lowest contour at 
2.5$\sigma$ level.}
\end{figure}

We show the $J$--$H$ and $H$--$K$ colour maps of NGC 6574 in Fig. 7. Most, 
but not all, \BG emission regions are situated close to the reddest colours. 
Both colour maps reveal the existence of a small spiral or bar structure 
around the nucleus. It is situated inside the spiral arms, but does not 
perfectly follow the morphology of the primary or nuclear stellar bar or the 
gaseous nuclear bar. In Fig. 9, we show the $J$--$K$ colour map, overlaid in 
black contours by the H$_2$ image. The H$_2$ emission follows the morphology 
in the $J$--$K$ colour map, and it is likely that the spiral/bar structure 
delineates the inflow route of the gas to the nucleus. 

\subsubsection{Star formation properties}

\begin{table*}
\begin{center}
\caption{Br$\gamma$ emission regions in NGC 6574}
\begin{tabular}{lccccccllcccccc}
  & & &\multicolumn{4}{c}{Observed emission}&&\multicolumn{5}{c}{Dereddened 
emission}\\
\hline
n & ap & corr$^a$ & Br$\gamma$ & H$_2$ & J--H & H--K & A$_V$$^b$ & Br$\gamma$ & H$_2$ & $\frac{H_2}{Br\gamma}$ & J--H & H--K \\
 & $''$ &  & \multicolumn{2}{c}{10$^{-15}$ ergs s$^{-1}$ cm$^{-2}$} & mag & 
mag & mag & \multicolumn{2}{c}{10$^{-15}$ ergs s$^{-1}$ cm$^{-2}$} & & mag & 
mag \\
 1& 6.7& 1.09& 3.40& 0.73& 0.76& 0.31& 0.95& 3.65& 0.79& 0.43&0.65& 0.27\\
 2& 1.7& 1.05& 0.19& 0.36& 0.76& 0.32& 0.86& 0.21& 0.38& 1.84&0.67& 0.28\\
 3& 1.7& 1.29& 0.24& 0.11& 0.75& 0.29& 3.2 & 0.31& 0.14& 0.47&0.44& 0.12\\
 4& 1.7& 1.35& 0.27& 0.16& 0.76& 0.31& 2.1 & 0.32& 0.19& 0.60&0.56& 0.20\\
 5& 3.4& 1.45& 1.94& 0.40& 0.77& 0.33& 2.7 & 2.37& 0.68& 0.21&0.50& 0.19\\
 6& 1.7& 1.64& 0.37& 0.39& 0.73& 0.27& 3.0 & 0.46& 0.49& 1.05&0.45& 0.12\\
 7& 1.7& 1.79& 0.31& 0.64& 0.78& 0.31& 3.5 & 0.40& 0.84& 2.08&0.45& 0.13\\
 8& 1.7& 1.76& 0.35& 0.07& 0.73& 0.29& 2.4 & 0.41& 0.08& 0.22&0.50& 0.17\\
 9& 3.4& 1.83& 1.03& 0.20& 0.71& 0.26& 1.6 & 1.16& 0.23& 0.19&0.55& 0.18\\
10& 2.2& 1.53& 0.48& --  & 0.78& 0.30& 1.8 & 0.55& --  & --  &0.59& 0.21\\
11& 2.8& 1.15& 0.80& 0.13& 0.76& 0.31& 0.51& 0.83& 0.14& 0.16&0.69& 0.30\\
nucl& 2.2& 1.01& 0.33& 0.51& 0.83& 0.31& 2.6 & 0.40& 0.62& 1.56&0.58& 0.18\\
\hline
\end{tabular}
\end{center}
$^a$: Correction factor for the velocity field and the change of wavelength 
across the array.\\
$^b$: Determined from the H$\alpha$/Br$\gamma$ ratio.\\
\end{table*}
\normalsize

The observed results for the \BG emission regions in NGC 6574 are given in 
Table 4. Note that the region marked with {\em nucl} is {\em close to} but 
not coincident with the nucleus. Using the H$\alpha$ image of NGC 6574 
(Gonz\'alez--Delgado et al. 1997) and the \BG image, we derived extinctions 
of A$_V$ = 0.51--3.5 (average A$_V$ = 2.1) for the emission regions. To our 
knowledge, there are no previous determinations of extinction for the ring of 
NGC 6574. The observed fluxes of the emission regions were corrected for the 
extinction, resulting in values listed in Table 4. 

The observed H$_2$/Br$\gamma$ ratios (0.16--2.1, average 0.73) are very 
similar to those found for NGC 1097 (Section 3.1.2). In most regions, the 
dominant excitation mechanism of the molecular gas is UV radiation from hot 
young stars (Puxley et al. 1990). Only in a few regions, the line ratio can 
be explained better by shocks. 

\begin{table*}
\begin{center}
\caption{Star formation properties of NGC 6574}
\begin{tabular}{lcccccccccr}
  & & &  \multicolumn{4}{c}{instantaneous star formation}  &
         \multicolumn{4}{c}{constant star formation rate}\\
\hline
n & N(H$^0$) & EW & age & mass & SFR$^a$ & v$_{SN}$ & age & mass$^b$ & SFR & v$_{SN}$ \\
 & 10$^{52}$ s$^{-1}$ & \AA & Myr & 10$^{6}$ M$_\odot$ & M$_\odot$ yr$^{-1}$ 
& 10$^{-3}$ yr$^{-1}$ & Myr & 10$^{6}$ M$_\odot$ & M$_\odot$ yr$^{-1}$ & 
10$^{-3}$ yr$^{-1}$\\ 
 1& 6.94& 22.54& 6.32& 18.79& 2.97& 19.3& 10.71& 8.20& 0.77& 3.2 \\
 2& 0.39& 14.81& 6.44& ~1.31& 0.20& ~1.2& 15.02& 0.64& 0.04& 0.3 \\
 3& 0.58& 37.36& 6.15& ~1.20& 0.20& ~1.1& ~9.01& 0.58& 0.07& 0.2 \\
 4& 0.60& 30.91& 6.22& ~1.37& 0.22& ~1.3& ~9.51& 0.63& 0.07& 0.2 \\
 5& 4.50& 72.24& 5.81& ~5.82& 1.00& ~4.9& ~7.93& 4.01& 0.51& 0.9 \\
 6& 0.88& 28.04& 6.26& ~2.14& 0.34& ~2.1& ~9.82& 0.96& 0.10& 3.3 \\
 7& 0.76& 59.36& 5.89& ~1.17& 0.20& ~1.0& ~8.21& 0.70& 0.09& 0.2 \\
 8& 0.79& 27.66& 6.26& ~1.92& 0.31& ~1.9& ~9.87& 0.86& 0.09& 0.3 \\
 9& 2.20& 22.12& 6.33& ~6.08& 0.96& ~5.8& 10.82& 2.63& 0.24& 1.0 \\
10& 1.04& 35.21& 6.18& ~2.24& 0.36& ~2.1& ~9.15& 1.06& 0.12& 0.4 \\
11& 1.58& 30.49& 6.23& ~3.67& 0.59& ~3.6& ~9.54& 1.67& 0.17& 0.6 \\
total  & 20.3  &       &      & 45.7  & 7.4 & 44.3 & & 21.9 & 2.3 & 10.6 \\
\hline
\end{tabular}
\end{center}
$^a$: Mass divided by age\\
$^b$: SFR multiplied by age\\
\end{table*}
\normalsize

The \BG EWs of NGC 6574 are small (15--72 \AA, average 35 \AA). Although the 
EWs are slightly larger than in NGC 1097, even for NGC 6574 a CSFR model with 
M$_u$ = 100 M$_\odot$ results in unreasonably large ages for the SF (up to 1 
Gyr; Table 5). These ages are much smaller (7.9--15 Myr) assuming the CSFR 
model with a reduced upper limit of the mass function (M$_u$ = 30 M$_\odot$). 
In this model, the SFR is 0.04--0.8 M$_\odot$ yr$^{-1}$ (average 0.21 
M$_\odot$ yr$^{-1}$; total within the ring 2.3 M$_\odot$ yr$^{-1}$) and the 
SN rate is 0.2--3.3 $\times 10^{-3}$ yr$^{-1}$ (average 1.0 $\times 10^{-3}$ 
yr$^{-1}$; total within the ring 1.1$\times 10^{-2}$ yr$^{-1}$). On the other 
hand, assuming an ISF model with M$_u$ = 100 M$_\odot$, the derived ages are 
even lower with a small spread, 5.8--6.4 Myr, very similar to those found for 
NGC 1097. The small range of these ages supports preferring the ISF model, 
but an unambiguous separation of the models is not possible with the current 
data. Assuming the ISF model, we derive for the masses of the \BG emission 
regions 1.2--19 $\times 10^6$ M$_\odot$ (average $4.2\times 10^6$ M$_\odot$, 
total within the ring $4.6\times 10^7$ M$_\odot$) and for the SN rate 
$v_{SN}$ 1.1--19 $\times 10^{-3}$ yr$^{-1}$ (average $4.0\times 10^{-3}$ 
yr$^{-1}$; total within the ring $4.4\times 10^{-2}$ yr$^{-1}$). Note that in 
Table 5, we do not list the values for the nucleus, since due to the weakness 
of the nuclear \BG emission, we could not derive a realistic EW for it after 
the subtraction of the bulge model. 

\begin{figure}
\psfig{file=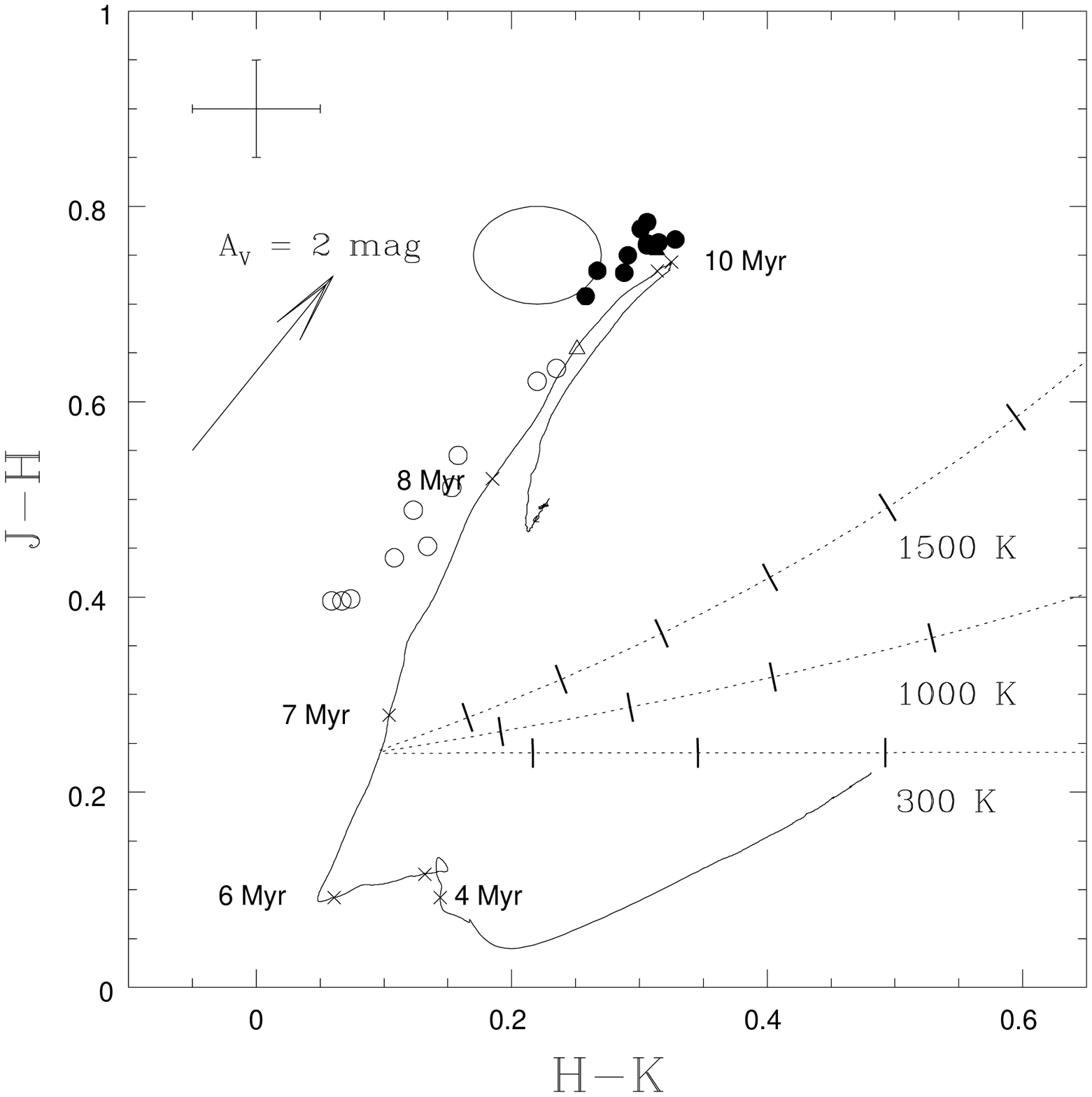,width=9cm,height=9cm}
\caption{As Fig. 5, but for NGC 6574.}
\end{figure}

Fig. 10 shows the $J$--$H$ and $H$--$K$ colour diagram of the \BG emission 
regions in NGC 6574. The observed $J$--$H$ colours are similar to normal 
unobscured spiral galaxies, but the $H$--$K$ colour is slightly redder, 
probably due to the contribution of hot dust emission. The 
extinction--corrected colours agree well with the L99 model but, as was found 
for NGC 1097, they indicate a slightly older population (7.5--8.5 Myr) than 
that derived from Br$\gamma$. A more physical model (e.g. a short duration or 
a decaying burst) would probably lead to a better agreement. 

\section{Discussion}

We have studied two active galaxies with circumnuclear SF. NGC 1097 has a 
clear SF ring, the diameter of which, 1.6 kpc, is typical for the nuclear 
rings surrounding galactic nuclei (Buta \& Crocker 1993). In NGC 6574, the 
\BG emission regions are located in spiral arms, forming a pseudoring. The 
morphology of the galaxies changes with wavelength. There is a generally 
good, but not perfect, agreement between the $K$--band and the \BG emission. 
The small differences probably reflect the existence of several SF episodes. 
Similarly, there is good but not perfect correlation between the H$_2$ and 
\BG morphologies. 

For both galaxies, we derived low \BG EWs in the cirumnuclear rings. 
Interpreting these values with a CSFR model and a high upper mass cutoff 
($M_u=100$ M$_\odot$) of the IMF leads to very old stellar population ages, 
up to 1 Gyr. Reducing the upper mass cutoff in the CSFR model to $M_u=30$ 
M$_\odot$ leads to more physically reasonable ages (9--50 Myr). The ISF model 
gives similar ages (6--7 Myr) for both galaxies, although there is a large 
scatter in the masses and SN rates of the emission regions. The small spread 
in the derived age both between the galaxies and within the emission regions 
is due to the very rapid decrease in the EW as a function of 
time in the L99 model at around 6--7 Myr, reflecting the fast evolution of 
the most massive stars away from the main sequence. It is possible that this 
similarity is a selection effect, as \BG can only reveal SF rings of a 
certain age. In NGC 1097, the observed radio spectral indices can not be 
explained in any CSFR model, so in this galaxy SF has already ended. A 
similar short duration (t $<$ 10$^6$ yr) SF episode has previously been 
suggested for e.g. the Seyfert 2 NGC 1068 (Davies et al. 1998), the starburst 
M82 (Rieke et al. 1993) and for the Galactic center (Genzel, Hollenbach \& 
Townes 1994). Further support for the short ages comes from the NIR 
broad--band morphology. The $K$--band morphology of NGC 1097 is clumpy, 
indicating relatively young age as the irregularities have not yet had time 
to smooth out, and suggesting the presence of RSGs. 

NGC 6574 has only a weak source of \BG emission in the nucleus, while no \BG 
was detected in the nucleus of NGC 1097. Either there is no SF in the 
nucleus, or it is too old to be detectable in Br$\gamma$. This situation 
agrees with the model of a propagating starburst, which has been suggested 
for e.g. Circinus (Maiolino et al. 1998), M82 (Satyapal et al. 1997) and IC 
342 (B\"oker, F\"orster--Screiber \& Genzel 1997). In this model, SF has 
either commenced from the nucleus, and propagated into the ring through 
shocks in the bar, or it has started at the ends of the bar and propagated 
elsewhere into the ring. Both galaxies have strong resolved nuclear H$_2$ 
emission. It is thus also conceivable that SF in the nucleus is prohibited 
and gas falls directly to the nucleus to fuel and obscure the AGN. Finally, a 
possible selection effect may explain the lack of observed SF in the nucleus, 
since if all rings are of roughly similar age, the nucleus may always be 
either too young or too old to produce a detectable amount of \BG emission. 

The observed H$_2$/Br$\gamma$ ratios in both galaxies show a large spread. 
The dominant excitation mechanism of the gas appears to be UV radiation from 
hot young stars. The largest ratios in NGC 1097 are found at the ends of the 
bar, indicating a contribution from shock excitation and supporting the model 
of a propagating starburst. 

The number of ionizing photons N(H$^0$) derived for NGC 1097 from thermal 
radio emission is much larger than that derived from \BG. This difference is 
unlikely to be caused by an incorrect determination of the radio spectral 
index, since the SN rate derived from non--thermal radio emission is in good 
agreement with that predicted by the L99 model. Either there exists thermal 
radio emission independently of SF, or the optical depth of \BG is large 
enough to allow the escape of only 10\%--20\% of the radiation from the 
emission regions. This would, however, indicate very large values of 
extinction A$_V$, which are not supported by the $JHK$ colours. 

Smith et al. (1999) proposed that the morphology of the circumnuclear ring 
reflects the mechanism which fuels the active nucleus. It is therefore 
instructive to compare the results found for NGC 1097 and NGC 6574 with other 
galaxies with circumnuclear rings with published \BG imaging. These include 
NGC 7771 (Reunanen et al. 1999), NGC 1808 (Kotilainen et al. 1996), NGC 7552 
(Schinnerer et al. 1997), NGC 1068 (Davies et al. 1998) and NGC 7469 (Genzel 
et al. 1995). There is no strong correlation between the ring morphology and 
the strength of the nucleus. The circumnuclear morphology of NGC 7771 is much 
clumpier than that of NGC 7552, while neither contains an AGN, but a nuclear 
starburst. NGC 1808 has a nuclear starburst with a possible weak Seyfert, and 
its morphology is rather smooth. The circumnuclear ring of NGC 1097 with a 
weak Seyfert/LINER nucleus is very clumpy, while the rings in the strong 
Seyferts NGC 6574 and NGC 7469 are rather smooth. Finally, NGC 1068 with a 
strong Seyfert nucleus has a clumpy ring morphology. 

A stronger inverse correlation is, on the other hand, found between the 
strength of the AGN and the strength of the bar. All three galaxies with a 
strong AGN (NGC 1068, NGC 6574 and NGC 7469) have a rather weak primary bar, 
whereas the four galaxies with a strong bar (NGC 1097, NGC 1808, NGC 7552 and 
NGC 7771) have either a weak or no AGN. This relationship agrees with the 
paucity of strong nuclear bars in AGN (Regan \& Mulchaey 1999) and argues 
against the $'$bars within bars$'$ scenario (Shlosman et al. 1989) as the 
primary fuelling mechanism of nuclear activity. However, Seyfert galaxies have 
large--scale bars more often than non--active galaxies (Knapen et al. 1999), 
therefore a much larger sample of barred galaxies with circumnuclear rings is 
needed to study the implications of this correlation for the fuelling of AGN.

\begin{acknowledgements} 
The United Kingdom Infrared Telescope is operated by the Joint Astronomy 
Centre on behalf of the U.K. Particle Physics and Astronomy Research Council. 
Thanks are due to Tom Geballe and Thor Wold for assistance during the 
observations, and to Rosa Gonz\'alez--Delgado for kindly making the NGC 6574 
H$\alpha$ image available to us. This research has made use of the NASA/IPAC 
Extragalactic Database (NED), which is operated by the Jet Propulsion 
Laboratory, California Institute of Technology, under contract with the 
National Aeronautics and Space Administration.\\
\end{acknowledgements} 

\noindent{\bf References}

\noindent Appleton P.N., Struck-Marcell C., 1996, Fund. Cosmic Phys. 16, 111\\
\noindent Barth A.J., Ho L.C., Filippenko A.V., Sargent W.L.W., 1995, AJ 110, 1009\\
\noindent Black J.H., van Dishoeck E.F., 1987, ApJ 322, 412\\
\noindent Bland--Hawthorn J., 1995, In: Comte G., Kylafis N. (eds.) ASP Conf. Ser. 71, Tridimensional optical spectroscopic methods in astrophysics, ASP, San Francisco, p. 72\\
\noindent Blitz L., 1991, In: Lada C., Kylafis N. (eds.) ASI Series 342, The Physics of Star Formation and Early Stellar Evolution, Kluwer Academic Publishers, Dordrecht, p. 3\\
\noindent Buta R., Crocker D.A., 1993, AJ 105, 1344\\
\noindent B\"oker T., F\"orster--Schreiber N.M., Genzel R., 1997, AJ 114, 1883\\
\noindent Calzetti D., 1997, AJ 113, 162\\
\noindent Cervino M., Mas--Hesse J.M., 1994, A\&A 284, 749\\
\noindent Charbonnel C., D\"{a}ppen W., Schaerer D. et al., 1999, A\&AS 135, 405\\
\noindent Combes F., Gerin M., 1985, A\&A 150, 327\\
\noindent Condon J.J., 1992, ARA\&A 30, 575\\
\noindent Condon J.J., Yin Q.F., 1990, ApJ 357, 97\\
\noindent Davies R.I., Sugai H., Ward M.J., 1998, MNRAS 300, 388\\
\noindent Demoulin M.--H., Chan Y., 1969, ApJ 156, 501\\
\noindent de Vaucouleurs G., de Vaucouleurs A., Corwin H.G. et al., 1991, 3rd Reference Catalogue of Bright Galaxies, Springer--Verlag\\
\noindent Devereux N.A., Kenney J.D., Young J.S., 1992, AJ 103, 784\\
\noindent Draine B.T., Roberge W.G., Dalgarno A., 1983, ApJ 264, 485\\
\noindent Elmegreen B.G., 1994, ApJ 425, L73\\
\noindent Evans I.N, Koratkar A.P., Storchi--Bergmann T. et al., 1996, ApJS 105, 93\\
\noindent Fernandez B.R., Holloway A.J., Meaburn J., Pedlar A., Mundell C.G., 1999, MNRAS 305, 319\\
\noindent Friedli D., Martinet E., 1993, A\&A 277, 27\\
\noindent Genzel R., Hollenbach D., Townes C.H., 1994, Rep. Prog. Phys. 57, 417\\
\noindent Genzel R., Weitzel L., Tacconi--Garman L.E. et al., 1995, ApJ 444, 129\\
\noindent Gerin M., Nakai N., Combes F., 1987, A\&A 203, 44\\
\noindent Glass I.S., Moorwood A.F.M., 1985, MNRAS 241, 429\\
\noindent Gonz\'alez--Delgado R.M., P\'erez E., Tadhunter C., Vilchez J.M., Rodr\'{\i}guez--Espinosa J.M., 1997, ApJS 108, 155\\
\noindent Hummel E., van der Hulst J.M., Keel W.C., 1987, A\&A 172, 32 (H87)\\
\noindent Knapen J.H., Shlosman I., Peletier R.F., 1999, ApJ, in press (astro-ph/9907379)\\
\noindent Kotilainen J.K., Forbes D.A., Moorwood A.F.M, van def Werf P.P, Ward M.J., 1996, A\&A 313, 771\\
\noindent Lancon A., Rocca--Volmerange B., 1996, New Astr. 1, 215\\
\noindent Landini M., Natta A., Oliva E., Salinari P., Moorwood A.F.M., 1984, A\&A 134, 284\\
\noindent Leitherer C., Schaerer D.,  Goldader J. et al., 1999, ApJS 123, 3 (L99)\\
\noindent Lejeune T., Buser R., Cuisinier F., 1997, A\&AS 125, 229\\
\noindent Maiolino R., Krabbe A., Thatte N., Genzel R., 1998, ApJ 493, 650\\
\noindent Maloney P.R., Hollenbach D.J., Tielens A.G.G.M., 1996, ApJ 466, 561\\
\noindent Mas--Hesse J.M., Kunth D., 1991, A\&AS 88, 399\\
\noindent Meaburn J., Fernandez B.R., Holloway A.J., et al., 1998, MNRAS 295, L45\\
\noindent Ondrechen M.P., van der Hulst J.M., 1983, ApJ 269, L47\\
\noindent Ondrechen M.P., van der Hulst J.M., Hummel E., 1989, ApJ 342, 39\\
\noindent Origlia L., Goldader J.D., Leitherer C., Schaerer D., Oliva E., 1999, ApJ 514, 96\\
\noindent Osterbrock D.E., 1989, Astrophysics of Gaseous Nebulae and Active Galactic Nuclei, University Science Books\\
\noindent P\'erez--Olea D.,E., Colina L., 1996, ApJ 468, 191\\
\noindent Pfenniger D., Norman C., 1990, ApJ 363, 391\\
\noindent Phillips M.M., Pagel,B.E., Edmunds,M.G., Diaz,A., 1984, MNRAS 210, 701\\
\noindent Piner B.G., Stone J.M, Teuben P., 1995, ApJ 449, 508\\
\noindent Puxley P.J., Hawarden T.G., Mountain C.M., 1990, ApJ 364, 77\\
\noindent Quillen A.C., Frogel J.A., Kuchinski L.E., Terndrup D.M., 1995, AJ 110, 156\\
\noindent Regan M.W., Mulchaey J.S., 1999, AJ 117, 2676\\
\noindent Reunanen J., Kotilainen J.K., Laine S., Ryder S.D., 1999, ApJ, in press (astro-ph/9909140)\\
\noindent Rieke G.H., Loken K., Rieke M.J., Tamblyn P., 1993, ApJ 412, 99\\
\noindent Ryder S.D., Staveley-Smith L., Malin D., Walsh W., 1995, AJ 109, 1592\\
\noindent Satyapal S., Watson D.M., Pipher J.L. et al., 1997, ApJ 483, 148\\
\noindent Schinnerer E., Eckart A., Quirrenbach A. et al., 1997, ApJ 488, 174\\
\noindent Shlosman I., Frank J., Begelman M.C., 1989, Nat 338, 45\\
\noindent Smith D.A., Herter T., Haynes M.P., Neff S.G., 1999, ApJ 510, 669\\
\noindent Storchi--Bergmann T., Baldwin J., Wilson A.S., 1993, ApJ 410, L11\\
\noindent Storchi--Bergmann T., Wilson A.S., Baldwin J., 1996, ApJ 460, 252\\
\noindent Telesco C.M., Dressel L.L., Wolstencroft R.D., 1993, ApJ 414, 120\\
\noindent Walborn N.R., Barba R.H., Brandner W. et al., 1999, AJ 117, 225\\
\noindent Walsh J.R., Nandy K., Thompson G.I., Meaburn J., 1986, MNRAS 220, 453\\

\end{document}